\documentclass[conference]{IEEEtran}
\makeatletter
\def\ps@headings{%
\def\@oddhead{\mbox{}\scriptsize\rightmark \hfil \thepage}%
\def\@evenhead{\scriptsize\thepage \hfil \leftmark\mbox{}}%
\def\@oddfoot{}%
\def\@evenfoot{}}
\makeatother
\usepackage{amssymb}
\usepackage{graphicx}
\usepackage{mediabb}
\usepackage{cite}
\usepackage{bm}

\def\bq{\begin{equation}}
\def\eq{\end{equation}}
\def\bqn{\begin{eqnarray}}
\def\eqn{\end{eqnarray}}
\def\bqnn{\begin{eqnarray*}}
\def\eqnn{\end{eqnarray*}}

\def\bfone{{\bf 1}}
\def\defeq{\stackrel{\mathrm{def}}{=}}

\newcommand{\lengthx}[1]{| #1 |_1}
\newcommand{\mod}[2]{\langle #1,#2 \rangle}
\newcommand{\modone}[1]{\langle #1 \rangle}

\newcommand{\senseResult}[3]{{#1}_d(#2,#3)}

\title{Estimating Target-Object Shape Using Location-Unknown Mobile Fixed-Direction Distance Sensors}
\author{
\IEEEauthorblockN{Hiroshi Saito}
\IEEEauthorblockA{NTT Network Technology Laboratory\\
3-9-11, Midori-cho, Musashino-shi, 180-8585 Tokyo, Japan\\
Email: saito.hiroshi@lab.ntt.co.jp}}

\date{}

\begin{document}

\maketitle
\begin{abstract}
This paper proposes a method of estimating a target-object shape, the location of which is unknown, through the use of location-unknown mobile distance sensors.
The direction of the sensor is fixed from the moving direction.
Typically, mobile sensors are mounted on vehicles. 
Each sensor continuously measures the distance from it to the target object.
The estimation method does not require any positioning function, anchor-location information, or additional mechanisms to obtain side information such as angle of arrival of signal. 
Under the assumption of a polygon target object, each edge length and vertex angle and their combinations are estimated to completely estimate the shape of the target object.

\end{abstract}

\section{Introduction}
Cars are being implemented with various distance sensors such as mm-wave sensors to prevent traffic accidents and improve the comfort of driving. Because some of these sensors have ranges larger than 100 meters, they can gather environment information. This environment information is used by the car itself and can be useful even for other cars or people. If such information is used by other people for other applications, this is vehicular-based participatory sensing or crowd sensing.

Although such an estimation intuitively seems impossible due to too many unknown factors and some theoretical results shown in the next section suggest it is impossible, this paper proposes a theoretical method for successfully estimating the target-object shape by using mobile sensors that continuously measure the distance between individual sensors and the target object.

\section{Related work}
The fundamental questions related to the research topic of this paper is whether we can estimate the shape of a target object using many simple sensors without a positioning function or location information and how we estimate it if possible. 
Studies by Saito et al. suggested that we can estimate only a small number of parameters such as the size and perimeter length of a target object with randomly deployed location-unknown simple sensors such as binary sensors and distance sensors and cannot estimate other parameters \cite{infocom,ieice-invite,arXiv}. 
Thus, they introduced composite sensors that are composed of several simple sensors and are randomly deployed.
By using them, additional parameters were able to be estimated \cite{signalProcess,mobileComp}.
The series of those studies used the sensing results at a certain sensing epoch and estimated parameters using them. 
Even when they used the sensing results at multiple sensing epochs, they did not take account of sensing epoch information. 
Only one study \cite{time-variant} in that series of studies took account of sensing epochs and the temporary behavior of sensing results, but it focused on estimating the size and perimeter length of the target object.

As far as we know, no studies other than the above series have directly tackled these questions. 
However, there have been considerable amount of studies on developing an estimation method that uses location-unknown sensors. These studies took a different approach. 
Most first estimated the sensor locations \cite{locating_nodes} because it is believed that ^^ ^^ the information gathered by such sensor nodes, in general, will be useless without determining the locations of these nodes" \cite{flip_amb} or ^^ ^^ the measurement data are meaningless without knowing the location from where the data are obtained" \cite{local_4}. 
Once sensors' locations are estimated, shape estimation is no longer difficult. 
However, an approach of estimating the sensor locations often requires additional mechanisms or side information, such as locations of anchor sensors and measurement mechanisms including angle-of-arrival measurements, training data and period, and distance-related measurements \cite{locating_nodes,local_4,local_2,local_3}. 
Concrete examples are intersensor distance information \cite{flip_amb}, location-known anchor sensors \cite{tsp2002}, a set of signals between sensors \cite{acm_sensor}, and the system dynamic model and location ambiguity of a small range \cite{bernoulli}.

In addition, there has been research into capturing the shape of a target object by using cameras that cannot cover the whole shape of the target object \cite{camera}.

\section{Model}
\begin{figure}[tb] 
\begin{center} 
\includegraphics[width=8cm,clip]{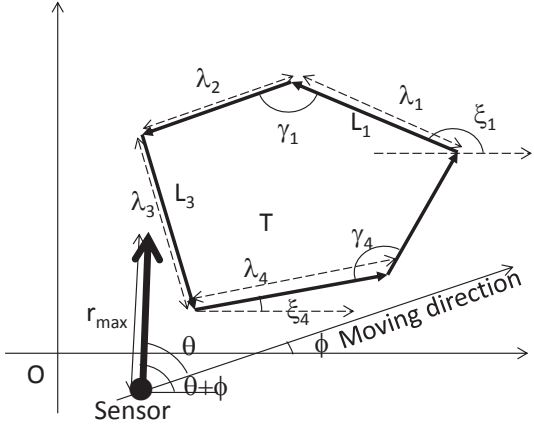} 
\caption{Illustration of target object model} 
\label{model} 
\end{center} 
\end{figure}

A target object $T$ is in a bounded convex set $\Omega\subset\mathbb{R}^2$.
It is a polygon, and its boundary $\partial T$ is closed and simple (no holes or double points) and consists of directional edges $\{L_j\}_j$ where $j =1,2,\cdots,n_e$ (Fig. \ref{model}).
Here, $n_e$ is the number of edges.
Let $\lambda_{j}$ be the length of $L_{j}$, and let $\xi_{j}$ be the angle formed by $L_{j}$ and the reference direction where $0\leq \xi_j<2\pi$.
Note that the inner angle formed by $L_{j}$ and $L_{j+1}$ is $\gamma_j=\pi-\xi_{j+1}+\xi_j$.
Here, $\{L_{j}\}_j$ are counted counterclockwise along $\partial T$ and the head of $L_{j}$ is the tail of $L_{j+1}$.
We do not know any of $\{\lambda_{j}, \xi_{j}, \gamma_j\}_{j}$.
That is, we do not know the target-object shape, size, or location.

A vehicle is running at a speed $v$ on a randomly placed straight line the direction of which is $\phi$ from the reference direction and passes through $\Omega$.
(This can be extended to a time-variant speed, but, for simplicity, assume that $v$ is time-invariant.)
It is equipped with a directional distance sensor the direction of which is $\theta$ from the moving direction.
(In practice, the vehicle's location (that is, the sensor's location) $(x_s(t),y_s(t))$ may not be in $T$, but the vehicle is assumed to run on a straight line passing through $T$ for simplicity.)
The sensor continuously measures the distance $r(t)\leq r_{max}$ at $t$ from the sensor to the target object and sends the sensing result.
Here, $r_{max}$ is the maximum range of the sensor.
Because the direction of the sensing range is $\phi+\theta$ from the reference direction, $r(t)$ is given as follows: 
$
r(t)=\cases{\tilde r(t), &if $\tilde r(t)\leq r_{max}$,\cr
\emptyset,&if $\tilde r(t)>r_{max}$.}
$
Here, $\tilde r(t)\defeq\min_{(x_s(t)+s\cos(\phi+\theta),y_s(t)+s\sin(\phi+\theta))\in T} s$.
In particular, $r(t)=0$ if $(x_s(t),y_s(t))\in T$.

The sensor continuously sends a report of $r(t)$ to an estimation server.
If $r(t)=\emptyset$, NO DETECTION is reported.
Although $\theta$ and $v$ are fixed and known, neither the vehicle's location $(x_s(t),y_s(t))$ nor moving direction $\phi$ is given to protect location privacy.
$\phi$ is a random variable uniformly distributed in $[0,2\pi)$.

There are $n_s$ vehicles monitoring $\Omega$, and each vehicle has a directional distance sensor.
$\theta,\phi,v,r(t)$ of the $i$-th vehicle or its sensor are described as $\theta_i,\phi_i,v_i,r_i(t)$.

\noindent
[Remark]
$T$ may consist of several polygons.
When obstacles are around a target object and sensors detect them, we should model the original target object and those obstacles as $T$.
As a result, we can estimate the shape of the original target object as well as those of obstacles.

Table \ref{p_list} lists the variables and parameters used in the remainder of this paper for the reader's convenience.
\begin{table}
\caption{List of variables and parameters}
\begin{center}\label{p_list}
\begin{tabular}{ll}
\hline
$T$&target object\\
$L_{j}$&$j$-th directional line segment of $\partial T$\\
$\lambda_{j}$&length of $L_{j}$\\
$\xi_{j}$&angle formed by $L_j$ and reference direction\\
$\gamma_j$&inner angle formed by $L_j$ and $L_{j+1}$\\
$n_e$&number of edges in $\partial T$\\
$n_s$& number of sensors\\
$r_{max}$&maximum sensing range\\
$\phi$&angle of vehicle's moving direction\\
$v$&moving speed of vehicle\\
$\theta$&sensing direction from vehicle's moving direction\\
$r(t)$&measured distance to $T$ at $t$\\
$p_d(L|\theta)$&period of $r(t)$ detecting a whole edge $L$ with direction $\theta$\\
$l_d(L|\theta)$&length in time of time period $p_d(L|\theta)$\\
$s_d(L|\theta)$&slope of $r(t)$ during $p_d(L|\theta)$\\
$\eta(\lambda,\theta)$&$\eta(\lambda,\theta)\defeq\arcsin\frac{r_{max}|\sin\theta|}{\lambda}$\\
$n_d(\lambda)$&number of sensors detecting the whole edge of length $\lambda$\\
$n_d(\gamma)$&number of sensors detecting a vertex of angle $\gamma$\\
$\Lambda_m$&$m$-th subset of temporary estimates of $\lambda$\\
$\Gamma_m$&$m$-th subset of temporary estimates of $\gamma$\\
\hline
\end{tabular}
\end{center}
\end{table}

In the remainder of this paper, we use the following notations: $\sharp(S)$ is the number of elements in a discrete set $S$, $\bfone(z)\defeq\cases{1, &if $z$ is true,\cr 0, &otherwise,}$, and $\widehat{z}$ is an estimator of $z$.
For angles $t_1,t_2$, $\modone {t_1}$ is $t_1$ under mod $2\pi$ and $\mod{t_1}{t_2}$ is an interval $[t_1,t_2)$ under mod $2\pi$.  
That is, $\mod{t_1}{t_2}$ is an interval $[t_1,t_2)$ if $t_1,t_2<2\pi$ and is intervals $[t_1,2\pi)\cup [0,t_2-2\pi)$ if $t_1<2\pi, 2\pi\leq t_2<4\pi$.

\section{Basic properties}\label{basic_pro}
\begin{figure}[tb] 
\begin{center} 
\includegraphics[width=8cm,clip]{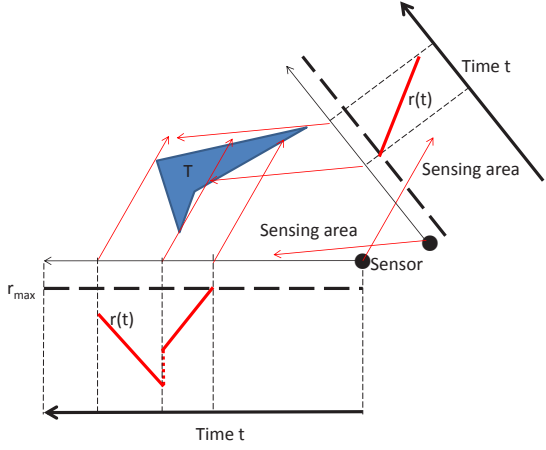} 
\caption{Basic example of $r(t)$} 
\label{basic-mobile} 
\end{center} 
\end{figure}

This section discusses basic properties of $r(t)$.
A simple example is illustrated in Fig. \ref{basic-mobile}.
An important observation of this figure is that there may be some jumps in $r(t)$ from a certain value between 0 and $r_{max}$ to another certain value.  
Only a single edge located nearest to a sensor is detected by the sensor, and its distance from the sensor is $r(t)$.
Even if other edges are within a sensing range, they are not detected or their distances to the sensor are not measured.
That is, detection of an edge may be blocked by another edge.
A jump down (up) of $r(t)$ occurs when a block starts (finishes).

\begin{figure}[tb] 
\begin{center} 
\includegraphics[width=8cm,clip]{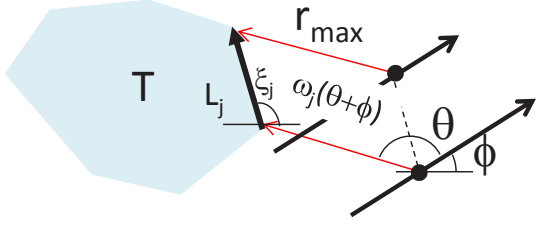} 
\caption{Illustration of $\omega_{j}(\theta+\phi)$} 
\label{omega} 
\end{center} 
\end{figure}

A sensor detects $L_{j}$ at $t$ if and only if the sensor is located in $\omega_{j}(\theta+\phi)$ at $t$, where $\omega_{j}(\theta+\phi)$ is a parallelogram attached to the right-hand side of $L_{j}$ and one of edges is $L_{j}$ and another has the length $r_{max}$ and the direction $\theta+\phi$ (Fig \ref{omega}). 
Note that the sensor detecting $L_j$ needs to satisfy 
\bq
\phi+\theta\in[\xi_j,\xi_j+\pi].\label{theta-xi}
\eq

\begin{figure}[tb] 
\begin{center} 
\includegraphics[width=8cm,clip]{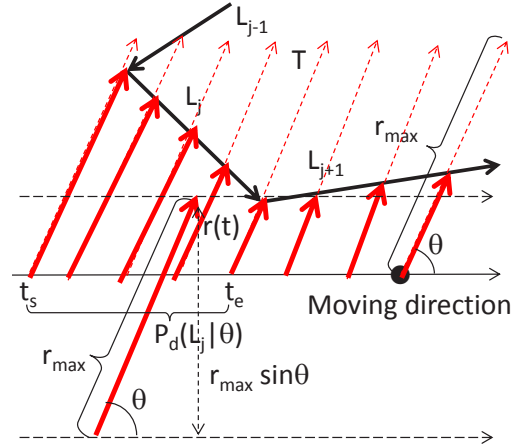} 
\caption{Case in which a sensor keeps detecting $L_{j}$} 
\label{single-edge} 
\end{center} 
\end{figure}

In the remainder of this section, we focus on the sensing results $r(t)>0$.
When a sensor keeps detecting an edge, $r(t)$ becomes continuous and becomes a line segment while the sensor keeps detecting it (Fig. \ref{single-edge}).
When the period detecting the {\it whole} $L_{j}$ with $r(t)>0$ by a sensor the direction of which is $\theta$ starts at $t_s$ and ends at $t_e$, an event corresponding to $t_s$ is (i) a change of slope at $r(t_s)>0$, (ii) a jump down of $r(t)$ at $t_s$, or (iii) $r(t_s)<r_{max}$ and $r(t_s-dt)=\emptyset$ and an event corresponding to $t_e$ is (i) a change of slope at $r(t_e)>0$, (ii) a jump up of $r(t)$ at $t_e$, or (iii) $r(t_e-dt)<r_{max}$ and $r(t_e)=\emptyset$.
(Note that the period does not include $r(t)=0$.)
In the remainder of this paper, we use a period $p_d(L|\theta)$ of $r(t)$ detecting a whole edge (that is, the period of $r(t)$ starting and ending with the events mentioned above) of $L$ with $r(t)>0$ for a given $\theta$ unless we explicitly indicated otherwise.
Let $l_d(L|\theta)$ and $s_d(L|\theta)$ be the length in time of $p_d(L|\theta)$ and the slope of $r(t)$ during $p_d(L|\theta)$.

For edge $L$ of length $\lambda$ and direction $\xi$ and the sensor the moving and sensing directions of which are $\phi$ and $\theta$, the following subsections provide (i) the relationships between the system parameters ($\lambda,\xi,\phi,\theta,v$) and the sensing results ($l_d(L|\theta),s_d(L|\theta)$), (ii) the probability that the sensor detects the whole $L$ with $r(t)>0$, and (iii) the probability that the sensor detects a vertex with $r(t)>0$.

\subsubsection{Relationships between system parameters and sensing results}
\begin{figure}[tb] 
\begin{center} 
\includegraphics[width=8cm,clip]{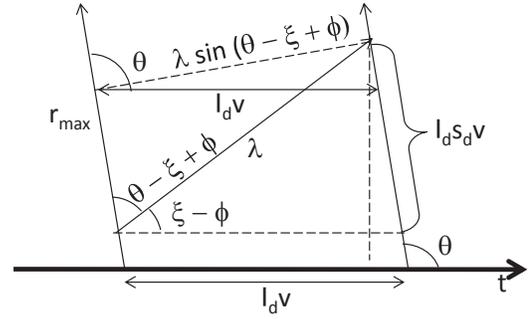} 
\caption{Illustration of $l_d$ and $s_d$} 
\label{ldsdFig} 
\end{center} 
\end{figure}

According to Fig. \ref{ldsdFig}, 
\bqn
vl_d(L|\theta)|\sin\theta|&=&\lambda\sin(\theta-\xi+\phi),\label{l_d}\\
vl_d(L|\theta)s_d(L|\theta)|\sin\theta|&=&\lambda\sin(\xi-\phi). \label{ldsd}
\eqn
Thus,
\bq
s_d(L|\theta)=-\sin(\xi-\phi)/\sin(\xi-\phi-\theta).\label{sd}
\eq

Because of Eq. (\ref{sd}), 
\bq
\xi-\phi=
\arctan{\frac{s_d(L|\theta)\sin\theta}{s_d(L|\theta)\cos\theta+1}}+(\pi)_{s_d(L|\theta)},\label{tan}
\eq
where $(\pi)_{s_d(L|\theta)}$ is 0 if $\modone {\arctan{\frac{s_d(L|\theta)\sin\theta}{s_d(L|\theta)\cos\theta+1}}-\theta}\in[\pi,2\pi)$ and is $\pi$ if $\modone {\arctan{\frac{s_d(L|\theta)\sin\theta}{s_d(L|\theta)\cos\theta+1}}-\theta}\in[0,\pi)$.
Apply this to Eq. (\ref{ldsd}) and obtain
\bq
\lambda=vl_d(L|\theta)\sqrt{s_d(L|\theta)^2+2s_d(L|\theta)\cos\theta+1}.\label{lam}
\eq


\subsubsection{Probability that sensor detects whole $L$ with $r(t)>0$}\label{sec-p_d}
\begin{figure}[tb] 
\begin{center} 
\includegraphics[width=8cm,clip]{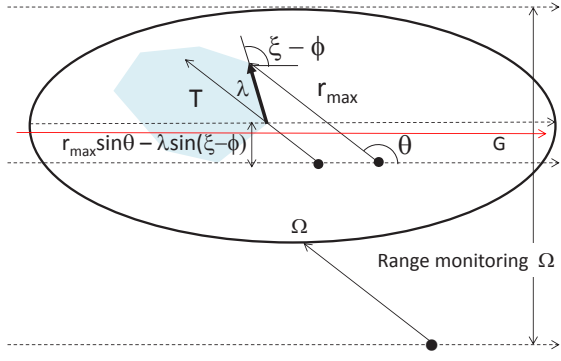} 
\caption{Location of sensors detecting whole edge with $r(t)>0$}
\label{omega2} 
\end{center} 
\end{figure}

According to Fig. \ref{omega2}, if line $G$ on which the sensor moves is in the directional strip of width $r_{max}|\sin\theta|-\lambda|\sin(\xi-\phi)|$ and if $\phi$ satisfies Eq. (\ref{theta-xi}) with $\xi_j=\xi$ (equivalently, $\phi-\xi\in [-\theta,-\theta+\pi]$), the sensor can detect the whole $L$.
Because the strip width must be non-negative, $\phi-\xi\in[-\eta,\eta]\cup[\pi-\eta,\pi+\eta]$ for $r_{max}|\sin\theta|<\lambda$ and $\phi-\xi\in[0,2\pi]$ for $r_{max}|\sin\theta|\geq\lambda$ where $\eta(\lambda,\theta)\defeq\arcsin\frac{r_{max}|\sin\theta|}{\lambda}\in [0,\pi/2]$.
(For simplicity, $\eta(\lambda,\theta)\defeq\pi/2$ for $r_{max}|\sin\theta|<\lambda$ in the remainder of this paper.)
Note that the measure of the set of $G$ on which sensors monitor $\Omega$ (Fig. \ref{omega2}) is given by Eq. (5.2) in \cite{santalo} and is $\lengthx{\Omega}+\pi r_{max}|\sin\theta|$.
Also note that the measure of the set of $G$ that is in this strip and has a direction satisfying Eq. (\ref{theta-xi}) is $\int_{\Phi_1}r_{max}|\sin\theta|-\lambda|\sin(\xi-\phi)|d\phi$ where $\Phi_1(\xi)\defeq ([\xi-\eta,\xi+\eta]\cup[\xi+\pi-\eta,\xi+\pi+\eta])\cap[\xi-\theta,\xi-\theta+\pi]$ for $r_{max}|\sin\theta|<\lambda$ and $\Phi_1(\xi)\defeq [\xi-\theta,\xi-\theta+\pi]$ for $r_{max}|\sin\theta|\geq\lambda$.
Because the probability $q_d(\lambda)$ that the sensor detects the whole $L$ of length $\lambda$ is given by the ratio of these measures in accordance with the definition of geometric probability \cite{santalo}, 
\bqn
q_d(\lambda)&=&\frac{\int_{\Phi_1}r_{max}|\sin\theta|-\lambda|\sin(\xi-\phi)|d\phi} {2\lengthx{\Omega}+2\pi r_{max}|\sin\theta|}\cr
&=&\frac{2\eta r_{max}|\sin\theta|-2\lambda(1-\cos\eta)}{2\lengthx{\Omega}+2\pi r_{max} |\sin\theta|}.\label{q_d_lam}
\eqn
(The denominator doubles because $G$ is directional.)
Therefore, the expected number $E[n_d(\lambda)]$ of sensors detecting the whole $L$ of length $\lambda$ with $r(t)>0$ is given by 
\bq
E[n_d(\lambda)]=\sum_{i=1}^{n_s}\frac{2\eta(\lambda,\theta_i) r_{max}|\sin\theta_i|-2\lambda(1-\cos\eta(\lambda,\theta_i))}{2\lengthx{\Omega}+2\pi r_{max} |\sin\theta_i|}.\label{num_detects1}
\eq

\subsubsection{Probability that sensor detects a vertex}
Here, we pay attention to the number of sensors that have sensing results that cover a vertex of $T$.
Such sensing results may not cover a whole edge.

Assume a vertex formed by $L_j,L_{j+1}$.
As shown in Fig. \ref{single-edge}, if line $G$ on which the sensor moves is in the directional strip of width $r_{max}|\sin\theta|$ and if $\phi$ and $\xi_j$ ($\xi_{j+1}$) satisfy Eq. (\ref{theta-xi}), the sensor can detect the part around the vertex with $r(t)>0$.
Similar to in Subsection \ref{sec-p_d}, the probability $q_d(\gamma_j)$ that the sensor can detect the vertex formed by $L_j,L_{j+1}$ with $r(t)>0$ is given by the following.
\bqn
&&q_d(\gamma_j)\cr
&=&\frac{\int_{[\xi_j-\theta,\xi_j-\theta+\pi]\cap[\xi_{j+1}-\theta,\xi_{j+1}-\theta+\pi]}r_{max}|\sin\theta|d\phi} {2\lengthx{\Omega}+2\pi r_{max}|\sin\theta|}\cr
&=&\cases{\frac{\gamma_j r_{max}|\sin\theta|}{2\lengthx{\Omega}+2\pi r_{max} |\sin\theta|},&for $\gamma_j\in(0,\pi)$,\cr
\frac{(2\pi-\gamma_j) r_{max}|\sin\theta|}{2\lengthx{\Omega}+2\pi r_{max} |\sin\theta|},&for $\gamma_j\in(\pi,2\pi)$,}\label{q_d_g}
\eqn
where $\gamma_j=\pi-\xi_{j+1}+\xi_j$ is the inner angle of the vertex.
Therefore, the expected number $E[n_d(\gamma)]$ of sensors detecting a vertex of inner angle $\gamma$ with $r(t)>0$ is given by 
\bq
E[n_d(\gamma)]=\cases{\sum_{i=1}^{n_s}\frac{\gamma r_{max}|\sin\theta_i|}{2\lengthx{\Omega}+2\pi r_{max} |\sin\theta_i|}, &for $\gamma\in(0,\pi)$,\cr
\sum_{i=1}^{n_s}\frac{(2\pi-\gamma) r_{max}|\sin\theta_i|}{2\lengthx{\Omega}+2\pi r_{max} |\sin\theta_i|}, &for $\gamma\in(\pi,2\pi)$.}\label{num_vertex}
\eq

\section{Estimation method}
Now, we are in a position to discuss target-object shape estimation.
The shape estimation method consists of four main parts and an additional part.
The first main part,  ^^ ^^ edge length estimation part," estimates the target object edge lengths $\{\lambda_j\}_j$.
The second main part, ^^ ^^ angle estimation part," estimates the angle of vertexes $\{\gamma_j\}_j$.
The third main part, ^^ ^^ combining length and direction estimation part," combines the results of the estimated edge lengths and angles and estimates a vertex formed by them.
The fourth main part, ^^ ^^ order estimation part," estimates of the order of the edges.
That is, it determines the consecutive edge of a certain edge.
Because we have already obtained the lengths and directions of edges at the end of the second main part, the estimated shape of $T$ is expected to be obtained at the end of the four main parts.
However, we need to compensate for estimation error when $T$ is not convex.
The additional part makes up for errors of estimating edges forming concave parts of $\partial T$.

\subsection{Edge length estimation part}
This part estimates edge length and the number of edges of the estimated edge length.
For a preliminary step, we need to obtain $(l_d,s_d)$ from the measured distance $r(t)>0$.
Assume that we obtain $\{(\senseResult l k i, \senseResult s k i)\}_k$ from the distance $r_i(t)$ measured by the $i$-th sensor where $\senseResult l k i$ ($\senseResult s k i$) is the $k$-th $l_d$ ($s_d$) derived from its sensing result observing a whole edge.
That is, when the $i$-th sensor observes $j$ whole edges of $T$, $r_i(t)$ has $j$ line segments corresponding to individual whole edges of $T$ and $\senseResult l k i$ and $\senseResult s k i$ are the length and slope of the $k$-th line segment among them.

For a given $(\senseResult l k i, \senseResult s k i)$ and known $v_i$ and $\theta_i$, obtain the temporary estimate of the edge length $\lambda$ due to Eq. (\ref{lam}).
\bq
\tilde{\lambda}(i,k)=v_i\senseResult l k i\sqrt{\senseResult s k i^2+2\senseResult s k i\cos\theta_i+1}.\label{est_lam0}
\eq
Intuitively, if the set $\{\tilde{\lambda}(i,k)\}_{i,k}$ forms $n_e$ clusters, each cluster corresponds to an edge.
To implement this intuition, classify the set of temporary edge length estimates.
It is a good idea to apply a classification tool such as Mclust of R \cite{mclust}.
Let $\Lambda_m$ be the $m$-th classified subset of this set of temporary estimates (or the set of $(l_d,s_d)$ deriving $\tilde{\lambda}(i,k)\in\Lambda_m$), and $n_\lambda$ be the number of the classified subsets (that is, $1\leq m\leq n_\lambda$).
By using $\tilde{\lambda}(i,k)\in\Lambda_m$, the mean of the temporary estimates in $\Lambda_m$ is adopted as the estimate of an edge length.
\bq
\widehat{\lambda}(\Lambda_m)=\sum_{\tilde{\lambda}(i,k)\in\Lambda_m}\tilde{\lambda}(i,k)/\sharp(\Lambda_m).\label{est_lam}
\eq

Here, note that $n_e=n_\lambda$ may not be valid.
This is because several edges may have the same length or classification may be incorrect.
To overcome this point, use Eq. (\ref{num_detects1}).
Note that the observed $n_d(\lambda)$ is $\sharp(\Lambda_m)$ when $\lambda=\tilde{\lambda}(i,k)$ for $\forall (i,k)\in\Lambda_m$, and that, if the number of edges of length $\lambda$ is $m$,  $E[n_d(\lambda)]$ is given by $m$ multiplied by the right-hand side of Eq. (\ref{num_detects1}).
Thus, the estimated number $\widehat{n_e}(\Lambda_m)$ of edges of length $\widehat{\lambda}(\Lambda_m)$ is shown below.
\bq
\widehat{n_e}(\Lambda_m)\approx\sharp(\Lambda_m)/E[n_d(\widehat{\lambda}(\Lambda_m))]\label{def_ne}
\eq


\subsection{Angle estimation part}
The angle estimation method proposed here applies Eq. (\ref{tan}) to two consecutive edges.
For a preliminary step, we need to find the sensing results $\{\senseResult s k i\}_{i,k}$ that cover a vertex.
Note that $r(t)$ is continuous when a sensor the direction of which is $\theta$ detects a vertex formed by $L_{j},L_{j+1}$.
Because $r(t)$ becomes a line segment for each edge, $r(t)$ becomes two consecutive line segments with different slopes for detected $L_{j},L_{j+1}$.
Thus, we apply Eq. (\ref{tan}) to $\senseResult s k i,\senseResult s {k+1} i$ corresponding to slopes of $r(t)$ detecting $L_{j},L_{j+1}$.
Note that we can use $\senseResult s k i,\senseResult s {k+1} i$ that cover only parts of the two edges $L_{j},L_{j+1}$ (not the whole $L_{j},L_{j+1}$).


By applying Eq. (\ref{tan}) to $\senseResult s k i,\senseResult s {k+1} i$, we can obtain the temporary estimate of the inner angle $\gamma_j$ formed by $L_{j},L_{j+1}$.
\bqn
\tilde \gamma_j(i,k)&=&\pi\pm(\arctan{\frac{\senseResult s k i\sin\theta_i}{\senseResult s k i\cos\theta_i+1}}+(\pi)_{\senseResult s k i}\cr
&&-\arctan{\frac{\senseResult s {k+1} i\sin\theta_i}{\senseResult s {k+1} i\cos\theta_i+1}}-(\pi)_{\senseResult s {k+1} i}),\cr
&&
\eqn
where $\pm$ becomes $+$ if $\sin\theta_i>0$ and becomes $-$ otherwise.
This is because $\senseResult s k i,\senseResult s {k+1} i$ corresponds to $L_{j},L_{j+1}$ if $\sin\theta_i>0$ and corresponds to $L_{j+1},L_{j}$ otherwise.

Similar to the edge length estimation, classify the set of temporary inner angle estimates.
Let $\Gamma_m$ be the $m$-th classified subset of this set of temporary estimates (or the set of measured slope pairs $\senseResult s k i,\senseResult s {k+1} i$ deriving $\tilde{\gamma}(i,k)\in\Gamma_m$), and let $n_\gamma$ be the number of the classified subsets (that is, $1\leq m\leq n_\gamma$).
By using $\tilde{\gamma}(i,k)\in\Gamma_m$, the estimate of an edge length is derived.
\bq
\widehat{\gamma}(\Gamma_m)=\sum_{\tilde{\gamma}(i,k)\in\Gamma_m}\tilde{\gamma}(i,k)/\sharp(\Gamma_m).\label{est_gam}
\eq

For each estimated $\widehat{\gamma}(\Gamma_m)$, there may be multiple vertexes.
Use Eq. (\ref{num_vertex}) to estimate the number of vertexes corresponding to the estimated $\gamma$.
Because the observed $n_d(\gamma)$ is $\sharp(\Gamma_m)$ when $\gamma=\tilde{\gamma}(i,k)$ for $\forall (i,k)\in\Gamma_m$, the estimated number $\widehat{n_e}(\Gamma_m)$ of vertexes that have inner angle $\widehat{\gamma}(\Gamma_m)$ is shown below.
\bq
\widehat{n_e}(\Gamma_m)\approx\sharp(\Gamma_m)/E[n_d(\widehat{\gamma}(\Gamma_m))]\label{angle_ne}
\eq

\subsection{Combining length and direction estimation part}
A vertex is determined by its inner angle and the lengths of two edges forming the vertex.
This part estimates the vertex by combining an estimated angle and estimated edge lengths obtained in the previous two parts.

If we can find such sensing results that $(\senseResult s k i,\senseResult s {k+1} i)\in \Gamma_m$, $(\senseResult l k i, \senseResult s k i)\in \Lambda_{m1}$, and $(\senseResult l {k+1} i, \senseResult s {k+1} i)\in \Lambda_{m2}$ and if $(\senseResult l k i, \senseResult s k i)$ and $(\senseResult l {k+1} i, \senseResult s {k+1} i)$ are sensing results of consecutive edges, these are sensing results of a vertex of angle $\widehat{\gamma}(\Gamma_m)$ and of lengths $\widehat{\lambda}(\Lambda_{m1})$ and $\widehat{\lambda}(\Lambda_{m2})$.
We count the number of such sensing results and judge that such a vertex exists if the counted number of results is large enough.

Although the angle estimation requires a pair of slope $\senseResult s k i,\senseResult s {k+1} i$ that may not cover the whole edges, the edge length estimation requires $\senseResult l k i, \senseResult s k i$ that covers the whole edge.
Therefore, it can happen that the angle of a vertex can be estimated but one of its edges (or any of its edges) cannot be estimated.
For the estimated angle $\widehat{\gamma}(\Gamma_m)$ and the edge length $\widehat{\lambda}(\Lambda_{m'})$, we count the number of sensing results satisfying $(\senseResult s k i,\senseResult s {k+1} i)\in \Gamma_m, (\senseResult l k i, \senseResult s k i)\in \Lambda_{m'}$ and judge the existence of the vertex of angle $\widehat{\gamma}(\Gamma_m)$ and of one of edge length $\widehat{\lambda}(\Lambda_{m'})$.

\subsection{Order estimation part}
To derive a method of identifying the order of edges, we use sensing results for consecutive edges.

Assume that $(\senseResult l k i, \senseResult s k i)\in \Lambda_m$ and $(\senseResult l {k+1} i, \senseResult s {k+1} i)\in\Lambda_{m'}$.
If a sensor continuously detects multiple edges without jumps of $r(t)$, they must be consecutive edges.
Therefore, an edge the length of which is estimated by $(\senseResult l k i, \senseResult s k i)$ likely connects to an edge the length of which is estimated by $(\senseResult l {k+1} i, \senseResult s {k+1} i)$.
Let $\Lambda(m,m')$ be the set of two sensing data pairs satisfying $\{(\senseResult l k i, \senseResult s k i)\in \Lambda_m,(\senseResult l {k+1} i, \senseResult s {k+1} i)\in \Lambda_{m'}\}_{i,k}$.
We judge that an edge of length $\widehat{\lambda}(\Lambda_{m})$ connects to an edge of length $\widehat{\lambda}(\Lambda_{m'})$, if $\sharp(\Lambda(m,m'))$ is large.

\subsection{Additional part}\label{sec-add}
This part may provide additional estimates of edges forming a concave vertex of $T$.
As described below, $\widehat{n_e}(\Lambda_m)$ defined by Eq. (\ref{def_ne}) may underestimate the number of edges for a non-convex $T$.
This part compensates for this error.

When $T$ has a concave part, $E[n_d(\lambda)]$ given by Eq. (\ref{num_detects1}) may not be correct.
The reason is as follows:
a sensor that should detect this edge may not do so because another edge of $\partial T$ is between this edge and this sensor.
That is, an edge of $\partial T$ blocks this sensor's detection of this edge.

\begin{figure}[tb] 
\begin{center} 
\includegraphics[width=8cm,clip]{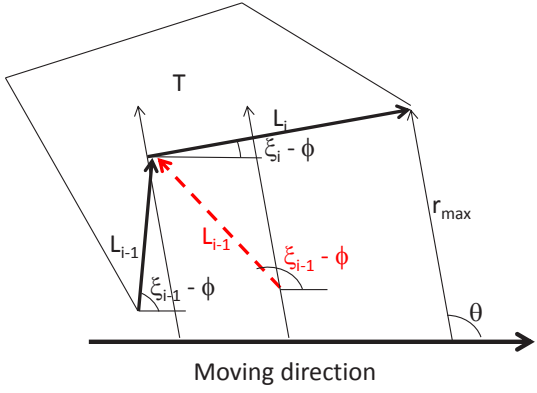} 
\caption{Blocking detection of $L_i$ due to convexity} 
\label{concave} 
\end{center} 
\end{figure}

We take account of this blocking and modify Eq. (\ref{num_detects1}) for a non-convex $T$.
We consider an event in which one consecutive edge $L_{i-1},L_i$ forming a concave vertex of $T$ may block the detection of the other edge.
We neglect other blocking events caused by other edges.
As shown in the derivation of Eq. (\ref{num_detects1}), $\phi\in \Phi_1(\xi_i)$.
This is because this sensor detects the whole $L_i$ and $r_{max}|\sin\theta|-\lambda_i|\sin(\xi_i-\phi)|>0$ (Fig. \ref{omega2}).
In addition, the detection of $L_i$ by the sensor the direction of which is $\theta$ is not blocked only if $\xi_{i-1}-\phi<\theta$ (Fig. \ref{concave}).
(Due to the concavity, $\delta\xi_{i-1}\defeq \gamma_{i-1}-\pi=\xi_{i-1}-\xi_i\in\mod 0 \pi$.)
Hence, the probability that a sensor detects the whole edge with length $\lambda_i$ and direction $\xi_i$ without blocking is given as follows.
\bq
q_d(\lambda_i,\theta,\delta\xi_{i-1})=\frac{f(\lambda_i,\theta,\delta\xi_{i-1})} {2\lengthx{\Omega}+2\pi r_{max} |\sin\theta|}\label{concave_f}
\eq
where $\phi\in\Phi_2(\xi_i,\xi_{i-1})\defeq \{\phi\in\Phi_1(\xi_1)\}\cap\{\phi>\xi_{i-1}-\theta\}$ and $f(\lambda_i,\theta,\delta\xi_{i-1})\defeq\int_{\phi\in\Phi_2(\xi_i,\xi_{i-1})} r_{max}|\sin\theta|-\lambda_i|\sin (\phi-\xi_i)|d\phi$ is given in Appendix \ref{app-add}.
The expected number $E[n_d(\lambda_i,\xi_i,\xi_{i-1})]$ of sensors detecting the whole $L_i$ of length $\lambda_i$ with $r(t)>0$ without blocking is given by
\bq
E[n_d(\lambda_i,\xi_i,\xi_{i-1})]=\sum_j q_d(\lambda_i,\theta_j,\delta\xi_{i-1}).
\eq

\section{Numerical examples}
In the remainder of this section, the following conditions are used as the default conditions unless explicitly indicated otherwise.
$\Omega$ is a disk with a radius of 200 length units.
$r_{max}=100$, $n_s=2000$. 
The sensing area direction is $\theta=\pi/2$ and the moving speed $v=1$ is for all the vehicles.

In the simulation conducted, each sensor sends a sensing report every single time unit.

\subsection{Basic properties}
\subsubsection{Impact of $\theta$}
\begin{figure}[thb] 
\begin{center} 
\includegraphics[width=8cm,clip]{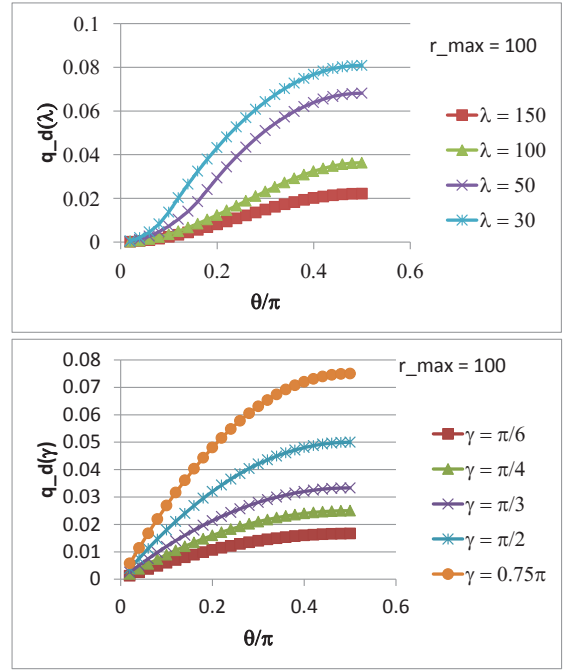} 
\caption{$q_d$ as function of $\theta$} 
\label{q_d} 
\end{center} 
\end{figure}

The proposed method uses the sensing data observing a whole edge or a vertex.
Therefore, the number of such data is very important for accurate estimation.
When $v$ and $\theta$ are the same for all the vehicles, the expected number of such data for a single edge or vertex is $E[n_d]=q_d n_s$ where $q_d$ is given by Eq. (\ref{q_d_lam}) or (\ref{q_d_g}).
Because $q_d$ is a function of $|\sin\theta|$, it is plotted against $\theta$ in Fig. \ref{q_d}.

As shown in this figure, $q_d(\lambda)$ and $q_d(\gamma)$ maximized at $\theta=\pi/2$.
This seems to be because the part a sensor detects while it moves becomes smaller as $|\sin\theta|$ becomes smaller.
For example, for $|\sin\theta|=0$, a sensor keeps detecting the same point of the target object and does not provide any information of other parts of the target object even though the vehicle moves.
Thus, small $|\sin\theta|$ results in a small amount of information that is useful in the proposed method.
In particular, we should avoid $|\sin\theta|\leq 0.5$ if possible.

For a fixed $\theta$, $q_d$ increases as an edge length becomes shorter or a vertex angle becomes wider.
This is because the shorter whole edge can be covered more easily than a longer one and because a wider angle can be detected more easily than a sharp one.

\subsubsection{Impact of $n_s$ and sensing error}
In addition to $\theta$, the number $n_s$ of sensors is a key parameter to determine the number of sensing data useful for the proposed method.
To evaluate the impact of $n_s$ on the estimation results, a simulation was conducted where $T$ is a right-triangle the edge lengths of which are 50, $50\sqrt{3}$, and 100.
Ten simulation runs were used for each value of $n_s$.
Furthermore, sensing errors were intentionally added.
Sensing errors for $s_d$ are a normally distributed random variable $\epsilon_s$ with mean 0 and standard deviation (s.d.) 0.03.
By adding an error, $s_d$ becomes $\tan(\arctan(s_d)+\epsilon_s)$.
A sensing report was lost with probability $\epsilon_l=0.002$.
As a result, $p_d(L|\theta)$ was divided at this epoch and $l_d(L|\theta)$ became shorter.

Two types of estimation errors occurred.
Type one is that the number of estimated edges and/or that of estimated angles became incorrect.
That is, the edge lengths (angles) derived by the proposed method were not three in the simulation for the right-triangle.
For example, the proposed method misjudged $T$ to have four edges.
Type two is that the estimated edge length or angle was inaccurate.

\begin{figure}[thb] 
\begin{center} 
\includegraphics[width=8cm,clip]{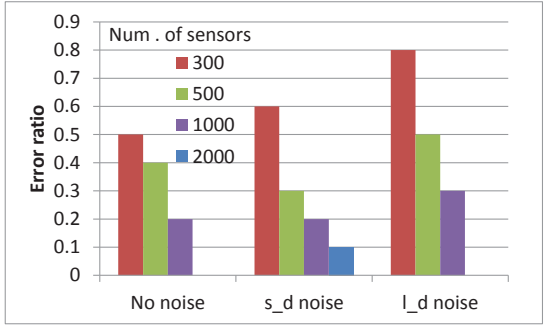} 
\caption{Error ratio} 
\label{error_ratio} 
\end{center} 
\end{figure}

\begin{figure}[thb] 
\begin{center} 
\includegraphics[width=8cm,clip]{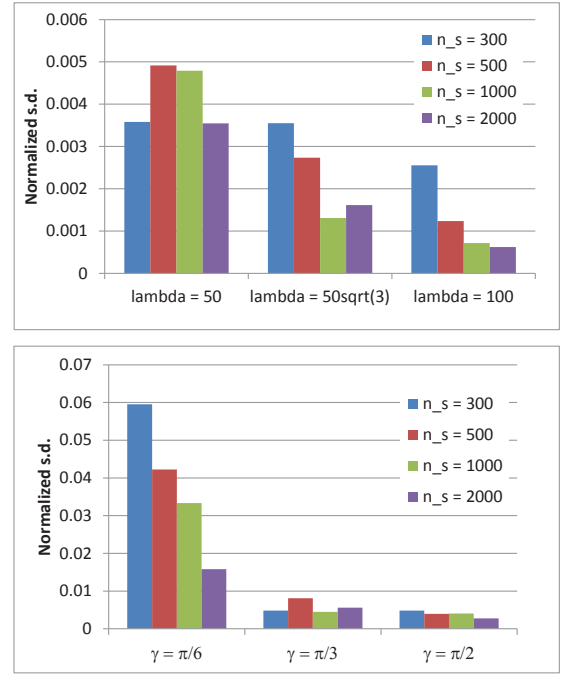} 
\caption{Normalized standard deviation of estimates (without sensing errors)} 
\label{estimation_accuracy} 
\end{center} 
\end{figure}

In the proposed method, $n_s$ has a large impact on type-one errors.
Figure \ref{error_ratio} plots the ratio of the number of simulation runs the results of which show type-one errors to the total number of simulation runs.
On the other hand, if there was no type-one error, the estimated edge lengths and angles were fairly accurate and insensitive to $n_s$ with/without sensing errors.
Figure \ref{estimation_accuracy} plots the standard deviation of $\widehat{\lambda}$ ($\widehat{\gamma}$) normalized by $\lambda$ ($\gamma$).
Although it decreased as $n_s$ became larger, it was small even for small $n_s$.
It was less than 1\% for edge length estimates and several\% or less for angle estimates.
This also shows that a longer edge (wider angle) has better estimation accuracy than a shorter (sharper) one.
(Although similar results were obtained with $\epsilon_s$ and $\epsilon_l$, they were omitted.) 
In addition, estimation bias was very small and fairly insensitive to $n_s$ with/without sensing errors.

As mentioned above, $n_s$ and $\theta$ have a large impact on the number of sensing data useful for the proposed method.
Therefore, $n_s$ also has a large impact on estimating the consecutive edges and the combination of an angle and edges forming a vertex, although no figures are shown.
As $n_s$ becomes smaller, the number of sensing results covering consecutive edges and those covering a vertex become smaller.
Thus, they become more difficult to estimate appropriately.

As noise ($\epsilon_l,\epsilon_s$) became larger, the estimation became less accurate.
Details are omitted due to the space limitations.

\subsection{Shape estimation of buildings}
\begin{figure}[tb] 
\begin{center} 
\includegraphics[width=8cm,clip]{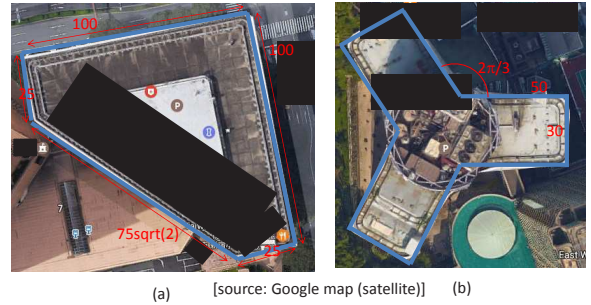} 
\caption{Examples of target objects} 
\label{building} 
\end{center} 
\end{figure}

The proposed method is applied to the buildings highlighted by thick blue lines in Fig. \ref{building}.
One is convex, and the other is concave.

\begin{figure}[tb] 
\begin{center} 
\includegraphics[width=8cm,clip]{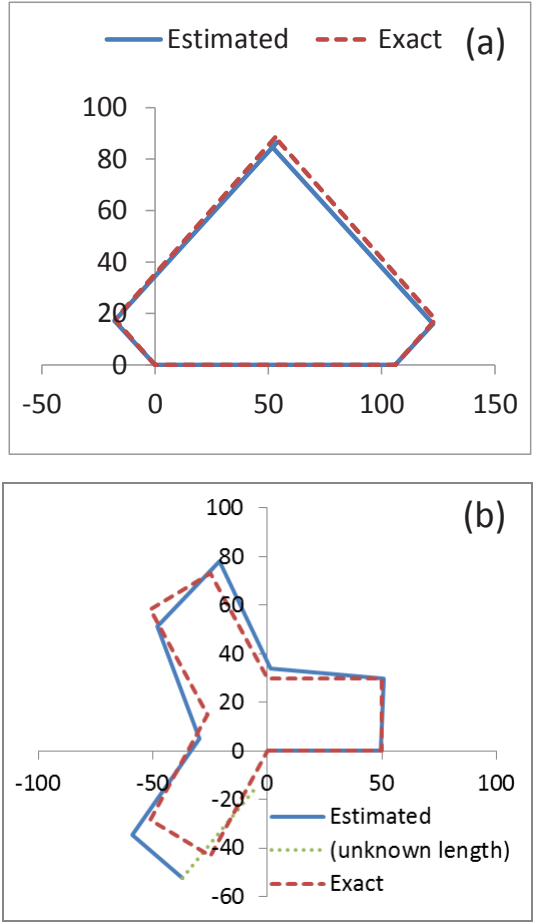} 
\caption{Estimated shape} 
\label{building-shape} 
\end{center} 
\end{figure}

For building (a), the estimated edge lengths and angles are shown in Table \ref{building-a}.
The estimation errors were several\% or less.
More than ten samples observed for the vertex of an estimated angle and two estimated lengths were $(\widehat{\gamma}(\Gamma_1),\widehat{\lambda}(\Lambda_2),\widehat{\lambda}(\Lambda_3))$, $(\widehat{\gamma}(\Gamma_1),\widehat{\lambda}(\Lambda_3),\widehat{\lambda}(\Lambda_2))$, $(\widehat{\gamma}(\Gamma_1),\widehat{\lambda}(\Lambda_3),\widehat{\lambda}(\Lambda_1))$, $(\widehat{\gamma}(\Gamma_1),\widehat{\lambda}(\Lambda_1),\widehat{\lambda}(\Lambda_3))$, $(\widehat{\gamma}(\Gamma_2),\widehat{\lambda}(\Lambda_1),\widehat{\lambda}(\Lambda_4))$, 
$(\widehat{\gamma}(\Gamma_2),\widehat{\lambda}(\Lambda_4),\widehat{\lambda}(\Lambda_1))$, 
$(\widehat{\gamma}(\Gamma_2),\widehat{\lambda}(\Lambda_2),\widehat{\lambda}(\Lambda_4))$, and $(\widehat{\gamma}(\Gamma_2),\widehat{\lambda}(\Lambda_4),\widehat{\lambda}(\Lambda_2))$.
Therefore, we can judge that a vertex of wide angle ($\widehat{\gamma}(\Gamma_2)$) is formed by the longest edge ($\widehat{\lambda}(\Lambda_4)$) and a short edge ($\widehat{\lambda}(\Lambda_1)$ or $\widehat{\lambda}(\Lambda_2)$).
Because it is estimated that there are two vertexes of wide angle and a single longest edge,  we can estimate that there are two of these vertexes and they are connected by the longest edge.
Let $L_2$ be the edge of length $\widehat{\lambda}(\Lambda_1)$.
Then, what we have estimated so far is: the estimated $\lambda_3$ and $\lambda_4$ are $\widehat{\lambda}(\Lambda_4)$ and $\widehat{\lambda}(\Lambda_2)$, and both the estimated $\gamma_2$ and $\gamma_3$ are $\widehat{\gamma}(\Gamma_2)$.
In addition, we can judge that a vertex of approximately $\pi/2$ ($\widehat{\gamma}(\Gamma_1)$) is formed by a short edge ($\widehat{\lambda}(\Lambda_1)$ or $\widehat{\lambda}(\Lambda_2)$) and a long edge ($\widehat{\lambda}(\Lambda_3)$).
Because there is a single edge of length $\widehat{\lambda}(\Lambda_1)$ and there is a single edge of length $\widehat{\lambda}(\Lambda_2)$, both the estimated $\lambda_1$ and $\lambda_5$ are $\widehat{\lambda}(\Lambda_3)$ and both the estimated $\gamma_1$ and $\gamma_4$ are $\widehat{\gamma}(\Gamma_1)$.
Thus, we can estimate the shape of this building even though there were not enough observed samples of a vertex formed by two long edges ($\widehat{\lambda}(\Lambda_3)$) or samples of consecutive long edges. 
The shape estimated is shown in Fig. \ref{building-shape}-(a).

\begin{table}
\caption{Estimated results for building (a)}
\begin{center}\label{building-a}
\begin{tabular}{lrrr}
\hline
&Estimated&Relative error&$\sharp$\\
\hline
$\widehat{\lambda}(\Lambda_1)$&23.17&-0.073&1\\
$\widehat{\lambda}(\Lambda_2)$&24.80&-0.008&1\\
$\widehat{\lambda}(\Lambda_3)$&99.77&-0.002&2\\
$\widehat{\lambda}(\Lambda_4)$&105.93&-0.003&1\\
$\widehat{\gamma}(\Gamma_1)$&1.610&0.025&3\\
$\widehat{\gamma}(\Gamma_2)$&2.379&0.010&2\\
\hline
\end{tabular}
\end{center}
\end{table}

For building (b), the estimated edge lengths and angles are shown in Table \ref{building-b}.
There are eight estimated edge lengths, although there are nine angles.
Such inconsistency can occur because the former needs the sensing data containing the whole edge and the latter needs the sensing data containing a vertex.
A wider angle generally has better estimation accuracy than a sharper one, but the concave angle had slightly poorer estimation accuracy than other convex angles in this example.

More than ten samples observed for the vertex of estimated angle and lengths are:
$(\widehat{\gamma}(\Gamma_1),\widehat{\lambda}(\Lambda_1),\widehat{\lambda}(\Lambda_3))$, $(\widehat{\gamma}(\Gamma_1),\widehat{\lambda}(\Lambda_3),\widehat{\lambda}(\Lambda_1))$, $(\widehat{\gamma}(\Gamma_1),\widehat{\lambda}(\Lambda_2),\widehat{\lambda}(\Lambda_3))$, $(\widehat{\gamma}(\Gamma_1),\widehat{\lambda}(\Lambda_3),\widehat{\lambda}(\Lambda_2))$, and 
$(\widehat{\gamma}(\Gamma_2),\widehat{\lambda}(\Lambda_3),\widehat{\lambda}(\Lambda_3))$.
Therefore, we judged that a concave vertex ($\widehat{\gamma}(\Gamma_2)$) is formed by two long edges ($\widehat{\lambda}(\Lambda_3)$) and that a vertex of approximately $\pi/2$  ($\widehat{\gamma}(\Gamma_1)$) is formed by the long edge  ($\widehat{\lambda}(\Lambda_3)$) and a short edge  ($\widehat{\lambda}(\Lambda_1)$ or  $\widehat{\lambda}(\Lambda_2)$).
The shape of the target object is shown in Fig. \ref{building-shape}-(b).

\begin{table}
\caption{Estimated results for building (b)}
\begin{center}\label{building-b}
\begin{tabular}{lrrr}
\hline
&Estimated&Relative error&$\sharp$\\
\hline
$\widehat{\lambda}(\Lambda_1)$&27.71&-0.076&1\\
$\widehat{\lambda}(\Lambda_2)$&29.71&-0.010&2\\
$\widehat{\lambda}(\Lambda_3)$&49.58&-0.008&5\\
$\widehat{\gamma}(\Gamma_1)$&1.615&0.028&6\\
$\widehat{\gamma}(\Gamma_2)$&4.156&-0.118&3\\
\hline
\end{tabular}
\end{center}
\end{table}

\section{Conclusion}
By using location-unknown distance sensors, this paper proposed a method of estimating the shape of a target object that is at  location is unknown.
Each sensor moves on an unknown line at a known speed and continuously measures the distance between it and the target object.
By collecting measured distances, the proposed method can estimate the shape of the target object.
The estimation method does not require any positioning function, anchor-location information, or additional mechanisms to obtain side information such as angle of arrival of signal. 

The successful development of this estimation method suggests that the possibility of software sensors implemented by participatory sensing under complete location privacy can be much wider.
It also suggests that the secondary use of IoT (internet of things) information can be wider than expected.

Because the estimation method assumed a polygon target object and a straight line trajectory of each sensor, the generalization of these assumptions remains as a further study.
An experiment using the proposed method is another future study.

\appendices

\section{Derivation of $f$ in Eq. \ref{concave_f}}\label{app-add}
To obtain $f$, use $x=\phi-\xi_i$ and integrate with $x$.

Assume $r_{max}|\sin\theta|<\lambda$.  
For $\eta\in [0,\pi/2]$, define $Z_1\defeq [0,\eta)$, $Z_2\defeq[\eta,\pi-\eta)$, $Z_3\defeq [\pi-\eta,\pi)$, $Z_4\defeq[\pi,\pi+\eta)$, $Z_5\defeq [\pi+\eta,2\pi-\eta)$, $Z_6\defeq[2\pi-\eta,2\pi)=[-\eta,0)$.  In the following, the operator $\in$ is applied in mod $2\pi$.

When $-\theta,\delta\xi-\theta\in Z_1$ or when $-\theta,\delta\xi-\theta\in Z_3$, $f=r_{max}|\sin\theta|\modone{2\eta-\delta\xi}-\lambda(2-2\cos\eta+\cos(\delta\xi-\theta)-\cos\theta)$.

When $-\theta\in Z_1,\delta\xi-\theta\in Z_2$, $f=r_{max}|\sin\theta|\modone{\eta-\theta}-\lambda(2-\cos\eta-\cos\theta)$.

When $-\theta\in Z_1,\delta\xi-\theta\in Z_3$, $f=r_{max}|\sin\theta|\modone{\pi-\delta\xi}-\lambda(2+\cos(\delta\xi-\theta)-\cos\theta)$.

When $-\theta\in Z_1,\delta\xi-\theta\in Z_4$ or when $-\theta\in Z_3,\delta\xi-\theta\in Z_6$, $f=r_{max}|\sin\theta|\modone{\pi-\delta\xi}-\lambda(-\cos(\delta\xi-\theta)-\cos\theta)$.

When $-\theta,\delta\xi-\theta\in Z_2$ or when $-\theta,\delta\xi-\theta\in Z_5$, $f=2\eta r_{max}|\sin\theta|-2\lambda(1-\cos\eta)$.

When $-\theta\in Z_2,\delta\xi-\theta\in Z_3$, $f=r_{max}|\sin\theta|\modone{\pi+\eta-\delta\xi+\theta}-\lambda(2+\cos(\delta\xi-\theta)-\cos\eta)$.

When $-\theta\in Z_2,\delta\xi-\theta\in Z_4$, $f=r_{max}|\sin\theta|\modone{\pi+\eta-\delta\xi+\theta}-\lambda(-\cos(\delta\xi-\theta)-\cos\eta)$.

When $-\theta\in Z_2,\delta\xi-\theta\in Z_5$ or when $-\theta\in Z_5,\delta\xi-\theta\in Z_2$, $f=0$.

When $-\theta\in Z_3,\delta\xi-\theta\in Z_4$, $f=r_{max}|\sin\theta|\modone{2\eta-\delta\xi}-\lambda(-2\cos\eta-\cos(\delta\xi-\theta)-\cos\theta)$.

When $-\theta\in Z_3,\delta\xi-\theta\in Z_5$, $f=r_{max}|\sin\theta|\modone{\eta-\theta-\pi}-\lambda(-\cos\eta-\cos\theta)$.

When $-\theta,\delta\xi-\theta\in Z_4$, $f=r_{max}|\sin\theta|\modone{2\eta-\delta\xi}-\lambda(2-2\cos\eta-\cos(\delta\xi-\theta)+\cos\theta)$.

When $-\theta\in Z_4,\delta\xi-\theta\in Z_5$, $f=r_{max}|\sin\theta|\modone{\pi+\eta-\theta}-\lambda(2-\cos\eta+\cos\theta)$.

When $-\theta\in Z_4,\delta\xi-\theta\in Z_6$,  $f=r_{max}|\sin\theta|\modone{\pi-\delta\xi}-\lambda(2-\cos(\delta\xi-\theta)+\cos\theta)$.

When $-\theta\in Z_4,\delta\xi-\theta\in Z_1$ or when $-\theta\in Z_6,\delta\xi-\theta\in Z_3$,  $f=r_{max}|\sin\theta|\modone{\pi-\delta\xi}-\lambda(\cos(\delta\xi-\theta)+\cos\theta)$.

When $-\theta\in Z_5,\delta\xi-\theta\in Z_6$,  $f=r_{max}|\sin\theta|\modone{\eta-\delta\xi+\theta}-\lambda(2-\cos(\delta\xi-\theta)-\cos\eta)$.

When $-\theta\in Z_5,\delta\xi-\theta\in Z_1$,  $f=r_{max}|\sin\theta|\modone{\eta-\delta\xi+\theta}-\lambda(\cos(\delta\xi-\theta)-\cos\eta)$.

When $-\theta,\delta\xi-\theta\in Z_6$, $f=r_{max}|\sin\theta|\modone{2\eta-\delta\xi}-\lambda(2-2\cos\eta-\cos(\delta\xi-\theta)+\cos\theta)$.

When $-\theta\in Z_6,\delta\xi-\theta\in Z_1$, $f=r_{max}|\sin\theta|\modone{2\eta-\delta\xi}-\lambda(-2\cos\eta+\cos(\delta\xi-\theta)+\cos\theta)$.

When $-\theta\in Z_6,\delta\xi-\theta\in Z_2$, $f=r_{max}|\sin\theta|\modone{\eta-\theta}-\lambda(-\cos\eta+\cos\theta)$.

Assume $r_{max}|\sin\theta|\geq\lambda$.  

When $-\theta\in [0,\pi),\delta\xi-\theta\in [0,\pi)$, $f=r_{max}|\sin\theta|\modone{\pi-\delta\xi}-\lambda(2+\cos(\delta\xi-\theta)-\cos\theta)$.

When $-\theta\in [0,\pi),\delta\xi-\theta\in [\pi,2\pi)$, $f=r_{max}|\sin\theta|\modone{\pi-\delta\xi}-\lambda(-\cos(\delta\xi-\theta)-\cos\theta)$.

When $-\theta\in [-\pi,0),\delta\xi-\theta\in [-\pi,0)$, $f=r_{max}|\sin\theta|\modone{\pi-\delta\xi}-\lambda(2-\cos(\delta\xi-\theta)+\cos\theta)$.

When $-\theta\in [-\pi,0),\delta\xi-\theta\in [0,\pi)$, $f=r_{max}|\sin\theta|\modone{\pi-\delta\xi}-\lambda(\cos(\delta\xi-\theta)+\cos\theta)$.


\begin{thebibliography}{99}
\bibitem{survey_privacy} J. Krumm, A survey of computational location privacy, Pers. Ubiquit. Comput., 13, pp. 391-399, 2009.
\bibitem{survey_privacy2} Mauro Conti, et al., Providing Source Location Privacy in Wireless Sensor Networks: A Survey, IEEE Communications Surveys \& Tutorials, 15, 3, pp.1238 - 1280, 2013.
\bibitem{IoT} J. Gubbi, et al., Internet of Things (IoT): A vision, architectural elements, and future directions, Future Generation Computer Systems, 29, 7, pp. 1645-1660, 2013.
\bibitem{IoTsurvey} L. D. Xu, et al., Internet of Things in Industries: A Survey, IEEE Trans. Industrial Informatics, 10, 4, pp. 2233-2243, 2014. 
\bibitem{commag} H. Saito, et al., Wide Area Ubiquitous Network: The Network Operator's View of a Sensor Network, IEEE Communications Magazine, 46, 12, pp. 112-120, 2008.
\bibitem{newIoTsurvey} Usman Raza, et al., Low Power Wide Area Networks: An Overview, IEEE Communications Surveys \& Tutorials, 19, 2, pp. 855-873, 2017.
\bibitem{infocom} H. Saito, et al., Shape Estimation Using Networked Binary Sensors, INFOCOM 2009.
\bibitem{ieice-invite} H. Saito, Local Information, Observable Parameters, and Global View, IEICE Trans. Communications, E96-B, 12, pp.3017-3027, 2013.
\bibitem{arXiv} H. Saito, et al., Geometric Analysis of Observability of Target Object Shape Using Location-Unknown Distance Sensors, arXiv.

\bibitem{signalProcess} H. Saito, et al., Stochastic Geometric Filter and Its Application to Shape Estimation for Target Objects, IEEE Trans. Signal Processing, 59, 10, , pp. 4971-4984, 2011.
\bibitem{mobileComp} H. Saito, et al., Estimating Parameters of Multiple Heterogeneous Target Objects Using Composite Sensor Nodes, IEEE Trans. Mobile Computing, 11, 1, pp. 125-138, 2012. 

\bibitem{time-variant} H. Saito, et al., Parameter Estimation Method for Time-variant Target Object Using Randomly Deployed Sensors and its Application to Participatory Sensing, IEEE Trans. Mobile Computing, 14, 6, pp. 1259-1271, 2015.

\bibitem{locating_nodes} N. Patwari, et al., Locating the nodes, IEEE Signal Processing Magazine, 22, 4, pp. 54-69, 2005.
\bibitem{flip_amb}A. A. Kannan, et al., Analysis of Flip Ambiguities for Robust Sensor Network Localization, IEEE Trans. Vehicular Technology, 59, 4, pp. 2057-2070, 2010.

\bibitem{local_4} G. Mao, et al., Wireless sensor network localization techniques, Computer Networks, 51, 10, pp. 2529-2553, 2007.
\bibitem{local_2}F. Gustafsson, et al., Mobile positioning using wireless networks: Possibilities and fundamental limitations based on available wireless network measurements, IEEE Signal Process. Mag., 22, 4, pp. 41-53, 2005.
\bibitem{local_3} A. Sayed, et al., Network-based wireless location: Challenges faced in developing techniques for accurate wireless location information, IEEE Signal Process. Mag., 22, 4, pp. 24-40, 2005.
\bibitem{tsp2002}J. C. Chen, et al., Maximum-Likelihood Source Localization and Unknown Sensor Location Estimation for Wideband Signals in the Near-Field, IEEE Trans. Signal Processing, 50, 8, pp. 1843-1854, 2002.
\bibitem{acm_sensor} X. Nguyen, et al., A Kernel-Based Learning Approach to Ad Hoc Sensor Network Localization, ACM Trans. Sensor Networks, 1, 1, pp. 134-152, 2005.
\bibitem{bernoulli}B. Ristic, et al., A Tutorial on Bernoulli Filters: Theory, Implementation and Applications, IEEE Trans. Signal Processing, 61, 13, pp. 3406-3430, 2013.

\bibitem{camera}Yibo Wu, et al., Photo Crowdsourcing for Area Coverage in Resource Constrained Environments, INFOCOM 2017.

\bibitem{santalo} L. A. Santal\'o, Integral Geometry and Geometric Probability, Second edition. Cambridge University Press, Cambridge, 2004.
\bibitem{mclust} [Available online]  [accessed on January 25, 2016] https://cran.r-project.org/web/packages/mclust/mclust.pdf


\end{thebibliography}
\end{document}